\documentclass[twocolumn,aps,pra,showpacs]{revtex4}
\usepackage{epsfig}
\usepackage{color}
\usepackage{multirow}
\usepackage[mathcal]{euscript}
\usepackage{amsmath,amssymb}
\begin{document}
\title{Entanglement-based continuous-variable quantum key distribution with multimode states and detectors}
\author{Vladyslav C. Usenko}
\affiliation{Department of Optics, Palack\' y University, 17. listopadu 1192/12,  771~46 Olomouc, Czech Republic}
\affiliation{Bogolyubov Institute for Theoretical Physics, Metrolohichna street 14-b, 03680 Kiev, Ukraine}
\author{Laszlo Ruppert}
\affiliation{Department of Optics, Palack\' y University, 17. listopadu 1192/12,  771~46 Olomouc, Czech Republic}
\author{Radim Filip}
\affiliation{Department of Optics, Palack\' y University, 17. listopadu 1192/12,  771~46 Olomouc, Czech Republic}
\date{\today}
\begin{abstract}
Secure quantum key distribution with multimode Gaussian entangled states and multimode homodyne detectors is proposed. In general the multimode character of both the sources of entanglement and the homodyne detectors can cause a security break even for a perfect channel when trusted parties are unaware of the detection structure. Taking into account the multimode structure and potential leakage of information from a homodyne detector reduces the loss of security to some extent. We suggest the symmetrization of the multimode sources of entanglement as an efficient method allowing us to fully recover the security irrespectively to multimode structure of the homodyne detectors. Further, we demonstrate that by increasing the number of the fluctuating but similar source modes the multimode protocol stabilizes the security of the quantum key distribution. The result opens the pathway towards quantum key distribution with multimode sources and detectors.
%the produced entanglement efficiently averages and therefore, 
\end{abstract}

\maketitle

\section{Introduction} 
Quantum physics, as fundamental physical theory, has many times
demonstrated its power as a very efficient tool to describe and predict
behavior of simple quantum systems. Pioneering development of quantum
optics describing both particle and wave quantum phenomena of few
individual photons or couple of optical modes combined with modern quantum
information theory changed  the view on quantum
physics during the last two decades \cite{Zeilinger2000,Kok2010}. A joint venture of these fields has produced a very interesting
application of the quantum key distribution (QKD), where
security is guaranteed by the laws of quantum physics \cite{Gisin2002,Scarani2009}. Beyond the practical
interest, the QKD brings also an operational view on many notions of quantum
physics of simple systems, such as quantum superposition, quantum
correlations, quantum nonlocality \cite{Peres1993}. 

Originally, the concept of quantum key
distribution was initiated from two very different points, the
Bennett-Brassard (BB84) protocol exploiting the superpositions of the single-photon states
\cite{BB84} and the Ekert protocol \cite{Ekert} based on the Bell-type quantum
correlations between two photons \cite{Bell}.
Recently, two orthogonal directions of quantum key distribution have
expanded shifting these original protocols into even more extreme positions.
The Ekert protocol has been developed into the device-independent form of the QKD \cite{Acin2007,Liu2013}.
On the other hand, the BB84 protocol has
been translated to the semiclassical QKD scenario. It uses only continuous-variable (CV)
squeezed \cite{Ralph1999,Cerf2001} and even coherent \cite{Grosshans2002,Silberhorn2002,Grosshans2003a} states and homodyne detectors so that explicit particle
properties of quantum states are not required to provide security. The
semiclassical nature of light still surprisingly allows transmission of secure
key rate for the large distances \cite{Jouguet2013}. Advantageously, the noise on the trusted sides is tolerable, allowing
%both arbitrary trusted preparation noise and detection noise are tolerable if the untrusted channel is only attenuating 
CV QKD with the noisy homodyne detection \cite{Patron2009} and even with the thermal states \cite{Filip2008,Usenko2010,Weed2010,Weed2012}. Recently, the
entanglement-based CV QKD with homodyne detection overcoming limits of the
coherent state CV QKD \cite{Usenko2011} has been experimentally verified \cite{Madsen2012}. The Gaussian
entanglement used in this advanced protocol is semiclassical since it still admits a local hidden
variable model based on positive Wigner functions \cite{Ou1992}. However, it provides quite robust security for a large distance.

One important and principal aspect of CV QKD has been always
omitted: quantum states and homodyne detectors can be
generally multimode \cite{Opatrny2002,Wasil2006,Christ2011,Hockel2011,Poly2012,Morin2013}. The single mode approximation is therefore only a very
strict assumption used in all the security analysis of continuous-variable
QKD and an impact on the multimode structure on the security of CV QKD has not been
analyzed yet, although it was studied for discrete-variable QKD \cite{Helwig2009}. In addition, the recently generated bright and heavily
multimode entangled beams \cite{Perina2007,Isk2009,Perina2012,Allevi2012,Agafonov2010,Isk2012} could have a potential to be positively used in the
semiclassical CV QKD to improve or stabilize its security. It is also suggested by their recent successful use in quantum imaging \cite{Brida2010, Brida2011, Lopaeva2013}. In this paper, we propose the multimode CV QKD taking fully into account the intrinsic
multimode structure of sources, channels, and detectors. We study in detail the impact of such structure
on security of CV QKD in the cases when trusted parties are aware or unaware of it. We observe the generally negative effect
of multimode structure on CV QKD in combination with imperfections of a multimode channel. We therefore suggest methods to
improve multimode CV QKD by proper mode selection at the detection stage and by
balancing the source, which can fully recover the security of the single-mode scenario. Finally, we show the
surprisingly positive effect of the increasing number of modes on the security of the multimode CV QKD protocol.

The paper is structured as follows: in Sec. \ref{multiprot} we describe the multimode homodyne detection, introduce the multimode CV QKD protocol, and state the assumptions being followed, then in Sec. \ref{security} we analyze the security limitations of the protocol imposed by the multimode detection structure. After that in Sec. \ref{improvements} we describe the possible methods to improve the security of the protocol, including the knowledge of detection structure, mode balancing in the detection, mode symmetrization in the source, key rate stabilization in the case of fluctuating modes, and then finish the paper with the concluding remarks.

\section{Multimode protocol}
\label{multiprot}
In our work we consider the entanglement-based Gaussian CV QKD protocol, where data are obtained by the trusted parties from the homodyne measurement of the Gaussian distributed continuous variables, namely quadratures of the electromagnetic field. Unlike the traditional approach, when the sources and the detectors are assumed to be single mode, we take into account the possible multimode structure of the homodyne detection. To model a multimode extension of the homodyne detector we consider a local oscillator consisting of
$N$ independent principally distinguishable orthogonal modes with strong classical amplitudes $|\alpha_i|\exp(i\theta)$, $i=1,\ldots,N$ and identical phase $\theta$. We suppose that two intensity detectors forming the homodyne detector are not able to distinguish the modes and therefore, the ideal balanced homodyne detection (with unit efficiency and no electronic noise) will produce a photocurrent $i^{(N)}_-=\sum_{i=1}^{N}|g_i\alpha_i|\tilde{X_i}(\theta)$, where $\tilde{X_i}(\theta)=a_i\exp(i\theta)+a_i^{\dagger}\exp(-i\theta)$ is the quadrature operator of a single mode $i$, $g_i$ are electronic gains of the detector for individual modes, and $[a_i,a_j^{\dagger}]=\delta_{ij}$. We use homodyne gains $G_i=|g_i\alpha_i|$ to simplify the notation. If the input of homodyne detector is blocked, the detector measures the $N$-mode vacuum state with variance $V^{(N)}_0=\sum_{i=1}^{N}G_i^2$. A standard approach in the CV experiments is to normalize $G_i$ by $\sqrt{V^{(N)}_0}$ to obtain the joint quadrature
\begin{equation}\label{joint}
X^{(N)}(\theta)=\frac{\sum_{i=1}^{N}G_i\tilde{X_i}(\theta)}{\sqrt{\sum_{n=1}^{N}G^2_n}}
\end{equation}
of the multimode light. The normalized coefficients $\lambda_i=G_i/\sqrt{\sum_{n=1}^{N}G^2_n}$ satisfy $\sum_{i=1}^N\lambda_i^2=1$.
We can therefore equivalently model any multimode homodyne detector as a linear optical network placed before the standard single-mode homodyne detector. This approach has been used in quantum optics to describe the multimode homodyne for measurement of the internal correlations of optical pulses \cite{Opatrny1997,McAlister1997,Fiurasek2001}.
The other unmeasured modes of that linear optical network are lost in the detector; they principally can either leave homodyne detection (untrusted detector) or they are
protected against any external access (trusted detector). The generic case is depicted in Fig.~1. If $G_i=G$, the detection of all the modes is balanced, the N-mode vacuum has variance $V^{(N)}_0=NG^2$ and joint normalized quadrature simply becomes $X^{(N)}(\theta)=\frac{\sum_{i=1}^{N}\tilde{X_i}(\theta)}{\sqrt{N}}$, corresponding to the symmetrical linear network before the single-mode homodyne.

\begin{figure}
\centerline{\psfig{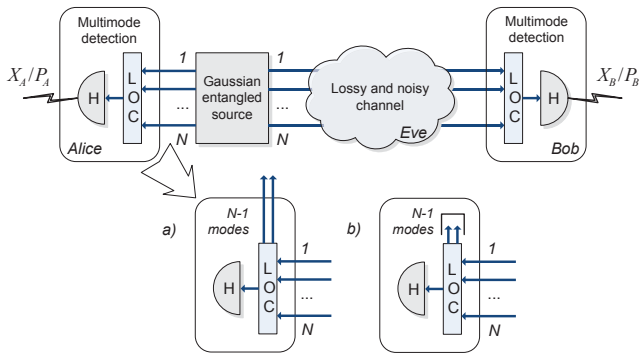}}
\caption{(Color online) Scheme of the multimode entanglement-based CV QKD with Gaussian entangled multimode source. The multimode
beams emitted by the source are measured by two trusted parties Alice and Bob after the untrusted lossy and noisy channel. The channel is
under control of an eavesdropper and is parametrized by transmittance $T$ and excess noise $\epsilon$. Measurement is modeled as performed by the single-mode homodyne detectors (H) precessed by the linear-optical modes coupling (LOC) and resulting in outcomes of the form (\ref{joint}). (a) Untrusted homodyne scenario, corresponding to the leakage of the auxiliary modes to the environment, where they can be potentially measured by an eavesdropper. (b) Trusted homodyne scenario, where auxiliary mode outputs of the LOC are discarded within the trusted stations.}
\label{scheme_general}
\end{figure}

Now let us consider the protocol based on the multimode homodyne detection and the generally multimode entangled states. The generic scheme of the multimode protocol is depicted in Fig. \ref{scheme_general}, where generally multimode entangled beams are measured by multimode homodyne detectors, modeled as single-mode homodyne detectors (H) precessed by linear-optical coupling of modes (LOC) and resulting in the outcomes of the form (\ref{joint}). The Gaussian distributed quadrature entangled multimode states produced by the source are defined by the variances $V_1\dots V_N$ in each of the $N$ measured modes. In this case, when a source does not emit in a particular i-th mode the variance of the respective vacuum state is thus $V_i=1$. Alternatively, when the source is multimode, it can be seen as a set of individual sources producing twin-beam states \cite{twinbeams} in the respective modes. The emitted modes are then measured by a multimode homodyne detector in one of the beams (A) at the sender trusted side (Alice). Another multimode beam travels through an untrusted channel, which is assumed to be fully controlled by a potential eavesdropper (Eve). The channel is parametrized by transmittance $T$ and excess noise $\epsilon$, which are assumed to be the same in all the modes (i.e., channel is mode insensitive). In practice this may not be always true (especially in the presence of chromatic dispersion and when the modes are detected in the long frequency range) but consideration of mode-sensitive channels will be the subject of future work. The channel then acts on a single-mode quadrature $\tilde{X_i}$ in the beam B so that it becomes $\sqrt{T}(\tilde{X_i}+X_N)+\sqrt{1-T}X_0$, where $X_N$ is the excess noise quadrature, having variance $\epsilon$, and $X_0$ is the multimode vacuum quadrature with variance 1, which is effectively coupled to the noisy signal on a beam-splitter $T$. A similar transformation follows then for the joint quadrature (\ref{joint}), measured by the multimode homodyne detector by the remote trusted party (Bob). 

In our work we waive the traditional assumption of the single-mode character of the signal. However, to be able to study the security of CV QKD in the multimode case and to clearly observe the multimode effects in CV QKD we state the following more detailed and thus weaker assumptions, which are related to the details of the multimode devices and which will be waived in further studies:
\begin{enumerate}
  \item No cross-talk is assumed between the modes, which are supposed to be statistically independent. 
  \item No mode mismatch is assumed at the multimode homodyne detection, i.e., all the modes are detected on both the detectors.
	\item The signal modes are assumed to be in phase with the respective local oscillator modes. In case this assumption does not hold, the additional detection noise would be present.
  \item The channel is assumed to be the same for all the modes.
	\item The multimode structure of the sources, channels, and detectors is completely known to Eve, but only the untrusted part of the setup can be manipulated by her. This is a typical assumption made in CV QKD.
\end{enumerate}
%%%%%%%%%%%%%%%%%%%%%%%%%%%%%%%%%%%%%%%%%%%%%%%%%%%%%%%%%%%%%%%%%%%%%%%%%%%%%%%%%%%%%%%%%%%
%%%%%%%%%%%%%%%%%%%%%%%%%%%%%%%%%%%%%%%%%%%%%%%%%%%%%%%%%%%%%%%%%%%%%%%%%%%%%%%%%%%%%%%%%%%%
%We assume typical conditions of
% either (A) prepare-and-measure CV QKD, when both state preparation at Alice's side and state measurement on Bob's are either fully trusted or untrusted, or (B) %%
%entanglement-based CV QKD, when the source is untrusted in the middle of the channel
%and both Alice's and Bob's detectors are either untrusted or trusted. 
The major point in the analysis of the multimode effects is the awareness of the trusted parties of the detection structure. If the trusted parties are unaware of the mentioned structure, all the residual modes in the multimode homodyne detection become attributed to Eve and contribute to her information on the key. If the trusted parties however know at least partially the detection structure, they can put a tighter bound on Eve's information.
%Eve's control of the auxiliary modes of the multimode homodyne detectors which are not measurable by Alice and Bob. If the multimode detectors are untrusted, although they are otherwise perfect, Eve has full access to the quantum states of these modes. If the multimode detector can be trusted, Eve would not have access to the auxiliary modes in the purification, as described in the next section.

\section{Security analysis}
\label{security}
The security analysis of the QKD protocols is generally based on the Csiszar-Korner theorem \cite{cktheorem}, from which follows that the secure key can be distilled from the classical information shared between the trusted parties if this information exceeds the information, available to a potential eavesdropper on the data possessed by either of the two trusted parties \cite{Gisin2002}. Thus, the security is established as the positivity of the lower bound on the secure key rate, which is the difference between the mutual information shared between the trusted parties, Alice and Bob, and the upper bound on the information, available to a potential eavesdropper, Eve. In the context of Gaussian CV QKD protocols the upper bound on Eve's information can be estimated depending on a particular class of attacks which Eve is assumed to be capable of. In the simpler case Eve is supposed to be able to individually measure her ancillary states coupled to the signal states, which is referred to as {\em individual attacks}. In the more general case of {\em collective attacks} it is assumed that Eve can store her ancillary states in a quantum memory and perform an optimized collective measurement on the set of states after the measurement bases are revealed by Alice and Bob. It was also recently shown that collective attacks are no less effective than the most general coherent attacks \cite{generalattacks1,generalattacks2}, which is also valid for continuous-variable protocols in the asymptotic regime \cite{cvcohatt}. Thus, we establish security against the individual attacks to analytically estimate the region where security is lost, since insecurity against individual attacks is a sufficient condition for insecurity of the protocol against more sophisticated attacks such as collective and coherent attacks. Alternatively we analyze security against collective attacks to estimate the region of parameters where the protocol is asymptotically secure against any type of attack. 

Mutual information shared between the trusted parties is estimated as Shannon (classical) mutual information. In the case when classical data possessed by the trusted parties is Gaussian distributed, the mutual information is defined by variances and conditional variances of the measurement outcomes and can be written as $I_{AB}=(1/2)\log_2{(V_A/V_{A|B})}$, where $V_A$ is the variance of the data measured by the multimode homodyne detector at Alice and $V_{A|B}=V_A-C_{AB}^2/V_{B}$ is the conditional variance of data measured at Alice on the measurement outcomes of multimode homodyne detection at Bob, defined through the correlation $C_{AB}$ between the measured data and the variance $V_{B}$ of the data measured at Bob. The explicit expressions for mutual information are given in the particular cases below. 

On the other hand, the estimation of Eve's knowledge on the measurement results of the trusted parties depends on the assumptions on the type of attack Eve is capable of and also on the knowledge the trusted parties have on the detection structure. We further distinguish the two cases: when the trusted parties are unaware of the detection structure and thus cannot distinguish whether the noise concerned with the detection is trusted (we refer to such a case as the {\em untrusted detection}); and when the trusted parties know the detection structure and are able to distinguish between the trusted detection noise and untrusted channel noise (further referred to as the {\em trusted detection}).

{\em Untrusted homodyne detectors.} For standard CV experiment with mode-insensitive multimode homodyne detectors and sources feeding these detectors by quantum multimode states, the trusted parties would typically not be able to recognize that the experiment actually deals with the multimode light. If the multimode detectors are untrusted, all the negative impact of multimode effects is attributed to the channel, i.e., Eve is capable of measuring the auxiliary modes leaving the linear-optical couplers [see Fig. \ref{scheme_general} (a)]. This case is equivalent to the untrusted preparation of a two-mode state, measured by the trusted parties, i.e., the entangled source is untrusted as well. In this case the channel becomes effectively present in both the beams of the entangled states and the protocol combines features of direct and reverse reconciliation schemes \cite{Grosshans2003a}, similarly to the case, when the source is placed in the middle of the channel \cite{Weed2013}. Let us consider this case in detail for both the individual and collective attacks.

{\em Security limitations.} We first study the case of the individual attacks which allows us to analytically identify the cases when the multimode CV QKD becomes insecure. The information, which is leaking and is thus potentially available to an eavesdropper Eve on the measurement results of the receiver trusted party Bob, in the case of individual attacks, is also given by the classical mutual information between the respective data, thus the lower bound on the key rate reads
\begin{equation}
\label{krind}
K_i=I_{AB}-I_{BE},
\end{equation}
where $I_{BE}=(1/2)\log_2{(V_B/V_{B|E})}$ and $V_{B|E}$ is the variance of the data measured by the remote trusted side (Bob) conditioned on the measurement outcomes of an eavesdropper (Eve).

The estimation of the second part of the lower bound (\ref{krind}), in particular, of conditional variance $V_{B|E}$ requires taking into account all the modes, which can be accessed by an eavesdropper for the measurement (i.e., all the untrusted modes). In the case of multimode detection this can be different depending on the knowledge of the trusted parties on the structure of their measurement. In the following we analyze security assuming two main scenarios—of the untrusted and trusted multimode homodyne detection. Note that mutual information $I_{AB}$ does not depend on whether we use a trusted or an untrusted detection model.

If the trusted parties are unaware of the multimode homodyne detection structure and only possess the data on the output in the form (\ref{joint}), the security analysis of the multimode CV QKD must take into account that the auxiliary output modes of the LOC are untrusted, i.e., available to an eavesdropper. 

To clearly demonstrate the multimode effect in the simplest case we first establish security bounds on the multimode protocol for the case of the individual attacks assuming that the channel is purely lossy, i.e., the channel excess noise $\epsilon=0$. Such a channel can be modeled as signal coupling to vacuum modes so that the eavesdropper can measure the respective output modes $E_{1...N}$ after the interaction. Moreover, if the multimode detection is untrusted, then the auxiliary output modes of the LOC interaction $A_{2...N}$ and $B_{2...N}$ on Alice and Bob trusted sides respectively (with no loss of generality we assume that mode with index 1 is measured on the output of LOC) are accessible to Eve. The conditional variance $V_{B|E}$ then becomes $V_{B_1|E_{1...N}A_{2...N}B_{2...N}}$. 

\begin{figure}
\centerline{\psfig{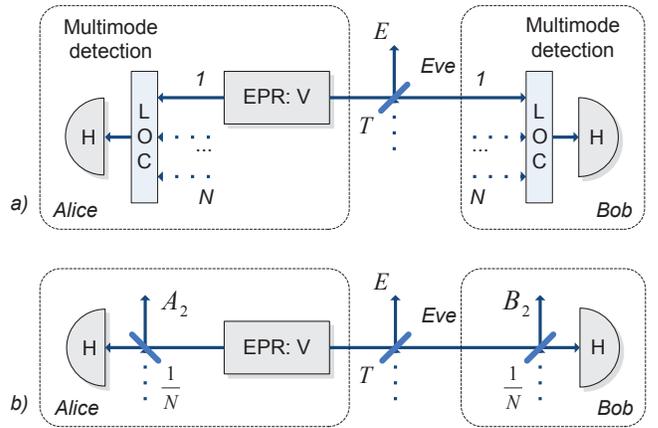}}
\caption{(Color online) Multimode CV QKD scheme with a single-mode source $V$ and $N$-mode detection in the general case (a) and the equivalent scheme in the case of balanced multimode detection (b). The particular case of the purely lossy channel (i.e., the excess noise $\epsilon=0$) is considered.}
\label{scheme_vacuum}
\end{figure}

Let us first consider the case when the source emits in a single mode, while the multimode detectors are measuring $N$ modes so that $N-1$ additional modes measured by the detectors are in the vacuum state, as depicted in Fig. \ref{scheme_vacuum} (a). In this case the multimode detection scheme becomes equivalent to coupling of a signal to vacuum with ratio $\lambda_1^2=1/N$ [see Fig. \ref{scheme_vacuum} (b)]. Such additional loss if it is attributed to the channel becomes equivalent to a side-channel leakage from both the entangled beams and may lead to a security break as well as to the reduction of entanglement shared between the trusted parties. 
%%%%%%%%%%%%%%%%%%%%%%%%%%%%%%%%%%%%%%%%%%%%%%%%%%%%%%%%%%%%%%%%%%%%%%%%%%%%%%%%%%%%%%%%%%%
%%%%%%%%%%%%%%%%%%%%%%%%%%%%%%%%%%%%%%%%%%%%%%%%%%%%%%%%%%%%%%%%%%%%%%%%%%%%%%%%%%%%%%%%%%%%
The mutual information in the case when all $N$ detected modes but one are in the vacuum state reads
\begin{equation}
I_{AB}=\frac{1}{2}\log_2{\frac{N+T(V-1)}{\frac{T(N-2)(V-1)}{N+V-1}+N}}
\end{equation}
In the case of the untrusted detection, the modes $A_2,B_2$ after coupling are available to an eavesdropper, so the conditional variance becomes $V_{B|E}=V_{B|EA_2B_2}$, where mode $B$ is measured by Bob and mode $E$ stands for the output of the purely attenuating channel (beamsplitter T). The lower bound on the information available to an eavesdropper then becomes
\begin{eqnarray}
\lefteqn{I_{BE}={}} \nonumber \\
& {}=\frac{1}{2}\log_2{\big[N+T(V-1)\big]\Bigg[\frac{T(N-2)(V-1)}{N^2(N+V-1)}+\frac{1}{N}\Bigg]}
\end{eqnarray}
The key rate (\ref{krind}) then turns to zero for any $T$ (even for $T=1$, i.e., the perfect untrusted channel) when $N=2$, i.e., the presence of one additional vacuum mode being measured already breaks the security of the protocol. This is in contrast to the single-mode protocols when any pure channel loss can in principle be tolerated with reverse reconciliation \cite{Grosshans2003a}. % Due to that the protocol becomes unsecure already in the case of the individual attacks, when the main channel is perfect and only single additional unoccupied mode is present. 
%The security break is caused by the untrusted loss of $50\%$ of the signal on the reference side of the protocol, which is the security bound for the direct reconciliation scheme in the single-mode regime. 
The security break is caused by Eve capturing $50\%$ of the signal on the side of Alice, which cancels Alice's advantage in knowledge of what was sent to the main channel and measured by Bob. Note that the security is lost although the Gaussian entanglement is preserved for this case. %This is also evident from the general considerations - the protocol in the case of untrusted detection consists of an untrusted channel in both the entangled beams and thus combines the case of reverse and direct reconciliation, since the channel is always present on the receiving side of the reconciliation. At the same time, it is known that the direct reconciliation scheme is not secure already for the channel transmittance of $50\%$, which is the case of $N=2$ in our scheme.

\begin{figure}
\centerline{\psfig{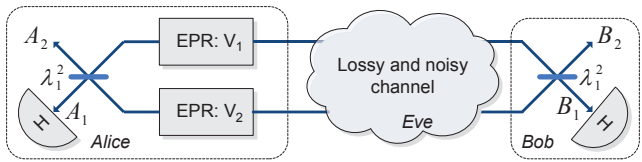}}
\caption{(Color online) Multimode CV QKD scheme with two-mode source and detection in the purification scheme, where modes are coupled and then measured by a single-mode homodyne detector.}
\label{scheme_two_mode}
\end{figure}

Now let us consider the case, when the source emits in two modes with state variances $V_1$ and $V_2$ in modes 1 and 2 respectively and only these modes are measured by the multimode detector. This is equivalent to mode coupling on a beamsplitter with transmittance $\lambda_1^2$ and measurement by a single-mode homodyne detector on one of the outputs (see Fig. \ref{scheme_two_mode}). In this case the mutual information between the trusted parties is given by
\begin{equation}
I_{AB}=\frac{1}{2}\log_2{
\frac{1}
{1-
\frac{T\Big(\lambda_1^2\sqrt{V_1^2-1}+(1-\lambda_1^2)\sqrt{V_2^2-1}\Big)^2}{A[1+T(A-1)]}
}
},
\end{equation}
where $A \equiv \lambda_1^2(V_1-V_2)+V_2$. 
%
%%%%%%%%%%%%%%%%%%%%%%%%%%%%%%%%%%%%%%%%%%%%%%%%%%%%%%%%%%%%%%%%%%%%%%%%%%%%%%%%%%%%%%%%%%%
%%%%%%%%%%%%%%%%%%%%%%%%%%%%%%%%%%%%%%%%%%%%%%%%%%%%%%%%%%%%%%%%%%%%%%%%%%%%%%%%%%%%%%%%%%%%
%
In the case of the pure channel loss and the untrusted detection, the upper bound on the leaking information is expressed through $V_B=T[\lambda_1^2V_1+(1-\lambda_1^2)V_2]+1-T$ and $V_{B|E_1E_2A_2B_2}$, which reads
\begin{widetext}
\begin{equation}
V_{B|E_1E_2A_2B_2}=\frac{1}{
1-\frac{T\Big(V_2-1+\lambda_1^2\big[2+V_1-V_2-2V_1V_2+2C_1C_2-2\lambda_1^2(1-V_1V_2+C_1C_2)\big]\Big)}
{\lambda_1^2V_1+(1-\lambda_1^2)V_2}
},
\end{equation}
\end{widetext}
where $E_{1,2}$ are the direct outputs of the purely lossy channels simulated by the beamsplitters (similarly to Fig. \ref{scheme_vacuum}); $A_2,B_2$ are the auxiliary output modes of the mode coupling (see Fig. \ref{scheme_two_mode}). From these expressions the security bounds can be obtained in the case of the individual attacks. 

{\em Applicability of the protocol.} Now let us analyze the security against the most general attacks in order to show if the multimode CV QKD is in principle possible. We consider the case when a potential eavesdropper is capable of collective attacks (equivalent to coherent attacks in the asymptotic regime), i.e., can store the ancillary states after coupling to a signal in a quantum memory and measure them collectively. In this case the upper bound on the information available to an eavesdropper on the measurement outcomes of the remote trusted party is given by the Holevo bound $\chi_{BE}$, which is the capacity of a bosonic channel between an eavesdropper and the reference side of the protocol (Bob in the case of the reverse reconciliation). The lower bound on the key rate then becomes
\begin{equation}
\label{krcol}
K=\beta I_{AB}-\chi_{BE},
\end{equation}
where we also took into account the effect of the imperfect postprocessing (mainly error correction) of the data, shared between the trusted parties, which is given by the postprocessing efficiency $\beta \in (0,1)$. The particular value of $\beta$ depends on the algorithms, used for the data processing, which also have different efficiencies depending on the signal-to-noise ratio, thus in the following analysis we check our results for typical values of $\beta$, which were about $95\%$ reached in the experiment with coherent-state CV QKD protocol \cite{lodewyck}, but can be also increased with the new effective error-correction codes \cite{effectivecodes}.

The Holevo bound is given by the von Neumann (quantum) entropies and in the case of the Gaussian distributed data reads $\chi_{BE}=S(\gamma_E)-S(\gamma_{E|B})$, where $S(\cdot)$ denotes the von Neumann entropy, $\gamma_E$ is the covariance matrix of a generally multimode Gaussian state available to an eavesdropper, and $\gamma_{E|B}$ is the covariance matrix of the state, which is available to an eavesdropper, conditioned by the measurement outcomes of the remote trusted party (Bob). The sufficiency of the covariance matrix description follows from the extremality of the Gaussian states \cite{extremality} and the subsequent optimality of the Gaussian collective attacks \cite{optimality1,optimality2}. 

The conditional covariance matrix after the measurement at the remote receiving party on mode $B$ is calculated as
\begin{equation}
\gamma_{E|B}=\gamma_E-\sigma_{EB}[X\gamma_B X]^{MP} \sigma_{EB}^T,
\end{equation}
where $\gamma_E$ is the covariance matrix of the generally multimode state which is available to an eavesdropper; $\gamma_B$ is the covariance matrix of the state of the mode $B$ prior to measurement; $\sigma_{EB}$ is the correlation matrix describing correlation between mode $B$ and the state given by covariance matrix $\gamma_E$; the diagonal $2x2$ matrix $X=Diag(1,0)$ stands with no loss of generality for the measurement of the $X$-quadrature in mode $B$; MP denotes Moore-Penrose pseudoinverse of a matrix.

The von Neumann entropy of a state given by an $N$-mode covariance matrix $\gamma$ is calculated from the symplectic eigenvalues  $\nu_i$ of the covariance matrix as 
\begin{equation}
S(\gamma)=\sum_{i=1}^N{G\Big(\frac{\nu_i-1}{2}\Big)},
\end{equation}
where $G[x]=(x+1)\log_2{(x+1)}-x\log_2{x}$ is the bosonic entropic function \cite{entrfunct}.

In the general case, when the channel is noisy or measurement is untrusted, the eavesdropper is assumed to be able to fully control the noise, i.e., hold the purification of the state, shared between the trusted parties. Then $S(\gamma_E)=S(\gamma_{AB})$, where $\gamma_{AB}$ is the covariance matrix of the state shared between Alice and Bob. Also, after the measurement in mode $B$ the similar equivalence holds for the conditional state: $S(\gamma_{E|B})=S(\gamma_{A|B})$. When multimode detection is untrusted, i.e., the trusted parties are unaware of the multimode structure of the their detectors and sources, only the two-mode matrix $\gamma_{AB}$ is available for the security analysis and represents a pessimistic (i.e., not tight) bound on the information, available to an eavesdropper. For ideal arbitrary asymmetrical detector and $N$ independent entangled state injected to the detector, the covariance matrix $\gamma^{(N)}_{AB}$ measured by Alice and Bob, being both "ignorant" about the multimode structure of the experiment, is equal to the weighted average
\begin{equation}
\gamma^{(N)}_{AB}=\sum_{i=1}^N \lambda^2_i \gamma_{AB,i}
\end{equation}
of the covariance matrices $\gamma_{AB,i}$ of individual modes. The covariance matrix $\gamma_{AB,i}$ is defined by $\gamma_{AB,i}^{(m,n)}=\langle r_i^{(m)} r_i^{(n)} \rangle - \langle r_i^{(m)}\rangle \langle r_i^{(n)}\rangle$ with $r_i:=(X_{A,i},X_{B,i},P_{A,i},P_{B,i})^T$ \cite{Weed2012b}. Such and "ignorant" approach is typical in the case when multimode structure is not taken into account. This approach may lead to the security break in terms of the key rate (\ref{krcol}) even for an ideal channel ($T=1,\epsilon=0$) and perfect postprocessing ($\beta=1$). 

In particular, when only one additional vacuum mode is present in the multimode detection (i.e., $N=2$, see Fig. \ref{scheme_vacuum}) and the noiseless channel is perfect ($T=1$), the mutual information becomes $I_{AB}=(1/2)\log_2{[(V+1)/2]}$, while the Holevo bound in this case is given by $G[(\nu-1)/2$, where $\nu=\sqrt{(V+1)/2}$, which evidently leads to security break.

When the two-mode source and two-mode detection is considered as given in Fig. \ref{scheme_two_mode}, the covariance matrix of the state, shared between the trusted parties, reads
\begin{widetext}
\begin{equation}
\gamma_{A_1B_1}=
\left( \begin{array}{cc}
[\lambda_1^2(V_1-V_2)+V_2]\mathbb{I} & \sqrt{T}[\lambda_1^2C_1+\lambda_2^2C_2]\sigma_z  \\
\sqrt{T}[\lambda_1^2C_1+\lambda_2^2C_2]\sigma_z & \{1+T[V_2+\lambda_1^2(V_1-V_2)-1]\}\mathbb{I}
\end{array} \right),
\end{equation}
\end{widetext}
where $C_{1,2} \equiv \sqrt{V^2_{1,2}-1}$, $\mathbb{I}=Diag(1,1)$ and $\sigma_z=Diag(1,-1)$. Similarly, the matrix of the uncorrelated two-mode output of the purely attenuating channel
\begin{equation}
\gamma_{E_1E_2}=
\left( \begin{array}{cc}
V_1(1-T)+T & 0 \\
0 & V_2(1-T)+T
\end{array} \right).
\end{equation}
The correlation matrix between the modes $A_1,B_1$ and $E_1,E_2$ reads
\begin{eqnarray}
\label{corrmat}
\lefteqn{\sigma_{(A_1B_1),(E_1E_2)}={}} \nonumber \\
& {}=-\left( \begin{array}{cc}
\sqrt{\lambda_1^2(1-T)}C_1\sigma_z & \sqrt{\lambda_1^2T(1-T)}C_1\sigma_z  \\
\sqrt{\lambda_2^2(1-T)}C_2\sigma_z & \sqrt{\lambda_2^2(1-T)T}C_2\sigma_z  
\end{array} \right).\;\;
\end{eqnarray}
The matrix of the auxiliary outputs of the LOC, modes $A_2,B_2$, which are available to the eavesdropper in the untrusted scenario, is the same as the matrix $\gamma_{A_1B_1}$ up to the substitution $V_1 \to V_2$ and vice versa. Finally, the correlation matrix $\sigma_{(A_1B_1),(A_2B_2)}$ between the trusted and untrusted modes of the two-mode homodyne detection reads
\begin{eqnarray}
\lefteqn{\sigma_{(A_1B_1),(A_2B_2)}={}} \nonumber \\
&{}=\left(\begin{array}{cc}
\lambda_1\lambda_2(V_2-V_1)\mathbb{I} & \lambda_1\lambda_2\sqrt{T}(C_2-C_1)\sigma_z  \\
\lambda_1\lambda_2\sqrt{T}(C_2-C_1)\sigma_z & T\lambda_1\lambda_2(V_2-V_1)\mathbb{I} 
\end{array}\right).\;\;\;\;
\end{eqnarray}
The results for the $N$ unoccupied auxiliary modes can be obtained from the covariance matrices given above by setting $V_2=1$ and $\lambda_1^2=1/N$.

%This loss caused by multimode detection can break the security completely, if it combined for example with the channel noise.

%(HERE, COMPLETE ANALYSIS OF ASYMMETRICAL TWO MODE CASE IN THIS APPROACH)

%{\bf It can introduce more loss and Gaussian noise into the multimode channel (??? more/less then to worst from the channels??, more general proof ??)}.

\section{Ways to improve security}
\label{improvements}
{\em Trusted homodyne detectors.} If we know, at least {\em a posteriori}, the exact structure of multimode homodyne detectors, multimode sources, and a channel, we can go beyond the "ignorant" untrusted approach and potentially enhance the security. We therefore consider the linear optical emulation (LOC) of the multimode detector as the purification of the otherwise ideal multimode detectors and consider un-measured outputs as trusted, i.e., not being accessible by Eve. This enables us to derive a tighter bound on Eve's information and increases the lower bound on the secure key rate thus improving the security of CV QKD. Note that the detectors must be in this case characterized independently of the protocol run since the parameters of the detection would be otherwise indistinguishable from the channel properties. A similar approach to the trusted detector characterization is used in the single-mode CV QKD \cite{Patron2009, lodewyck}. In the multimode case the detectors must be calibrated apart of the protocol using a known and controllable multimode source (studied with a multimode measurement \cite{Opatrny2002,Wasil2006,Christ2011,Hockel2011,Poly2012,Morin2013}) that allows obtaining the parameters of the LOC of the given multimode detector.  

However, the multimode influence in the case of the trusted homodyne detection becomes equivalent to the trusted preparation noise \cite{Filip2008,Usenko2010} and trusted detection noise \cite{Patron2009} for a single-mode protocol. This may still lead to security break, when auxiliary modes are not occupied.

\begin{figure}
\centerline{\psfig{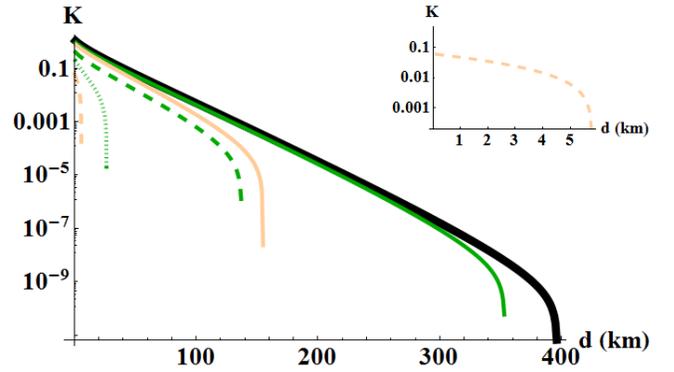}}
\caption{(Color online) Lower bound on secure key rate in the case of collective attacks versus distance in fiber with standard attenuation -0.2 dB/km for multimode entanglement-based CV QKD with two-mode states upon trusted, given in green (dark gray), and untrusted, given in orange (light gray), homodyne detection at $V_1=3$. $\lambda_1^2=0.5$ and $V_2=1$ (dotted lines),
$\lambda_1^2=0.5$ and $V_2=1.1$ (dashed lines), $\lambda_1^2=0.95$  and $V_2=1$ (solid lines). Black solid thick line represents the benchmark set by single-mode CV QKD protocol with source variance $V=3$. In all the cases channel noise is $\epsilon=5\%$ SNU and postprocessing efficiency $\beta=95\%$.}
\label{ked}
\end{figure}

The effect of trusted detection can be observed already in the case of individual attacks. The conditional variance in this case is obtained only from $V_{B|E}$ as an eavesdropper could not benefit from the measurement of the auxiliary modes. The second part of the key rate then reads 
\begin{equation}
I_{BE}=\frac{1}{2}\log_2{\frac{[N+T(V-1)][T(V-1)-V]}{T(N-1)(V-1)-NV}},
\end{equation}
and the protocol becomes secure for $N=2$ at any channel transmittance and is secure upon any $N$ when the channel is perfect (at least in the case of individual attacks; the collective attacks are considered below). However, the security break may still occur. In particular, the security bound is given by
\begin{equation}
\frac{1}{N}=\frac{(1-T)(V-1)}{3V-2T(V-1)-1},
\end{equation}
from which follows that for strongly attenuating channel $T\to 0$ the bound becomes $1/N=(V-1)/(3V-1)$ and for $V \to \infty$ the security break is observed when more than two additional vacuum modes are measured by the multimode homodyne detector.
%, and detection is strongly unbalanced towards the unoccupied mode, i.e., $\lambda_1 \to 0$. In such the case the security is lost upon $\lambda_1^2=(1-T)(V_1-1)/[3V_1-2T(V_1-1)-1]$.
%If the variance of the main signal mode is arbitrarily strong, i.e., $V_1 \to \infty$, the security is lost at $\lambda_1^2=(1-T)/(3-2T)$, which sets the bound of $\lambda_1^2=1/3$ for the strongly %attenuating channel $T \to 0$. Thus, no more than two additional vacuum modes can be tolerated in this case, when detection is balanced. The bound on detector balancing quickly saturates with the %variance of the main signal mode.
In the case of collective attacks the purification involves not only the modes $A_1$ and $B_1$, which are detected after LOC, but also all the auxiliary outputs of LOC, i.e., the equations $S(\gamma_E)=S(\gamma_{A_1...A_NB_1...B_N})$ and $S(\gamma_{E|B})=S(\gamma_{A_1...A_NB_2...B_N|B_1})$ hold. For the previously discussed basic two-mode scenario with one unoccupied mode we reach no security break when detection is balanced. In particular, for the arbitrarily strong variance of the main signal $V_1 \to \infty$ the secure key rate in the case of collective attacks reads $K^{(2)}=\frac{1}{2}\log_2{[(1-T/2)/(1-T)]}$, which is lower than the similar key rate in the single-mode case $K^{(1)}=\log_2{[1/(1-T)]}$, but in contrast to the untrusted case remains positive at any non-zero channel transmittance T. The improvement from trusted consideration of the multimode homodyne detectors can be also seen from the Fig. \ref{ked}. It shows the key rate upon the presence of additional unoccupied mode and trusted homodyne detector (green dotted line), which is non-zero contrary to the untrusted case, when security was lost already for a perfect channel. Here and further in our analysis we fix the channel noise to $5\%$ of the shot-noise unit (SNU) which well complies with the long-distance realization of coherent-state CV QKD \cite{Jouguet2013}, where the channel noise of up to $1\%$ SNU was observed.
%%%%%%%%%%%%%%%%%%%%%%%%%%%%%%%%%%%%%%%%%%%%%%%%%%%%%%%%%%%%%%%%%%%%%%%%%%%%%%%%%%%%%%%%%%%
%%%%%%%%%%%%%%%%%%%%%%%%%%%%%%%%%%%%%%%%%%%%%%%%%%%%%%%%%%%%%%%%%%%%%%%%%%%%%%%%%%%%%%%%%%%%
%e.g. being 0.3 already for $V_1=10$ and $T=3\cdot 10^{-2}$. The knowledge of the multimode structure may also be incomplete and thus lead to a partial improvement of the security region of the protocol (see Supplemental Material \cite{supp} for the discussion).

\begin{figure}[h]
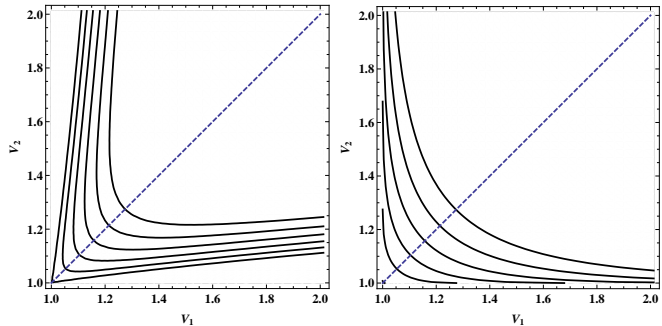

\begin{tabular}{ll}
\includegraphics[width=0.24\textwidth]{contour_un.eps}
\includegraphics[width=0.24\textwidth]{contour_tr.eps}	
\end{tabular}
\caption{(Color online) Security region against collective attacks for the two-mode CV QKD in terms of mode variances $V_1$ and $V_2$ in case of untrusted (left) and trusted (right) multimode homodyne detection. In both the cases channel noise $\epsilon=0,1,2,3,4,5\%$ SNU (lines from left to right) is present, mode coupling in detection is balanced. Channel transmittance $T=3\cdot 10^{-2}$, postprocessing is perfect ($\beta=1$). The area of parameters above each of the lines is secure upon given conditions. 
\label{contour}}
\end{figure}

{\em Unbalanced multimode sources.} Contrary to the case, when the auxiliary modes are not occupied, the source may emit in all or some of the modes measured by the multimode detector. The presence of the less occupied modes with the lower variance still limits security of CV QKD, as seen from Fig. \ref{contour}, which is further enforced by the presence of the channel noise. Evidently, the untrusted homodyne detection scenario is more sensitive to the unbalancing of the source. In particular, in the case of individual attacks and arbitrarily strong variance of the main mode $V_1 \to \infty$ the security break is observed independently of channel transmittance at any $V_2<1-\lambda_1^2+1/(4-4\lambda_1^2)$ if $\lambda_1^2\in(0,0.5)$, when the homodyne detection is untrusted. The negative impact of the unbalancing of the source is also visible in Fig. \ref{ked} in both the trusted and untrusted regimes. However, even the weak signal in the auxiliary mode can still improve robustness of the protocol; see Fig. \ref{ked} (dashed lines). 
%In particular, the security bound on $\lambda_1^2$ reaches 0.22 already for $V_2=1.01$ in the example given above when $V_1=10$. 
%%%%%%%
%%%%%%%

{\em Mode selection in the homodyne detector.} If the weighting in the detector enforces the mode with low correlation in the source, the impact on the resulting key rate is negative, as mentioned above, thus the security of the scheme is reduced. However, if the detection is mode-selective towards the modes with the high correlation in the source, the resulting key rate and security of the protocol can be restored. In the limiting case the proper detection can completely cancel the noisy uncorrelated modes. If the mode selection is imperfect, then it still limits the security region of the protocol, but the improvement can be significant, which is seen from Fig. \ref{ked} (solid lines).

\begin{figure}
\centerline{\psfig{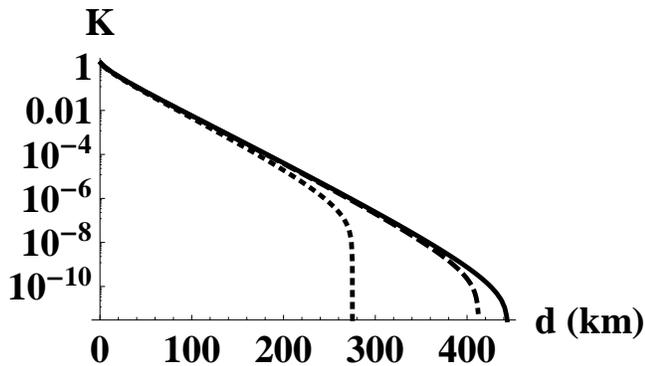}}
\caption{Lower bound on secure key rate in the case of collective attacks versus distance in fiber with standard attenuation -0.2 dB/km for multimode entanglement-based CV QKD upon three different situations given in Table \ref{equivtable}: 3-mode (solid line), 2-mode (dashed line), and 1-mode (dotted line); the postprocessing efficiency is $\beta=95\%$.}
\label{3scen}
\end{figure}

{\em Partial knowledge of the detection structure.} %We further consider the more general case of the collective attacks, which allows us to show the positive effects of the multimode structure in certain settings and also generalize the results for the %many-mode case.
%In the case when the trusted parties are aware of the multimode structure of their detector, i.e., the detection is trusted,  The calculations in this case as well as in the case of the noisy channels and %untrusted detection are done numerically with the results presented in Fig. 2,3,4 of the main manuscript.
It is also possible that the trusted parties possess only limited knowledge of the detection and source multimode structure, i.e., they can only partly purify the covariance matrix obtained from their measurement. Indeed, if the trusted parties are able to discriminate between some of the modes in the multimode detection, but are still unaware about the whole structure, the security bound is improved compared to a completely "ignorant" approach, but is still reduced compared to the complete knowledge of the setup structure. Moreover, the effective parameters of the untrusted channel can be recalculated depending on the ability of the trusted parties to discriminate between the modes and learn their structure. The example of the impact of partial knowledge of the trusted parties on the parameters of the setup is given in Table \ref{equivtable}, where different parametrizations of the channel leading to the same two-mode covariance matrix $\gamma_{AB}$ measured by the multimode homodyne detectors are given depending on how much information on the source and detection structure trusted parties have.
\begin{widetext}
\begin{center}
\begin{table}[h]
\begin{tabular}{c|c|c|c|}
\cline{2-4}
& 3-mode & 2-mode & 1-mode \\
& (reality) & (limited knowledge) & ("ignorant" approach) \\ \cline{1-4}
\multicolumn{1}{|c|}{} & $V_1=5$, $\lambda_1^2=95\%$ & $V_1^{(2)}=5$, $\lambda_1^2=95\%$ & \\
\multicolumn{1}{|c|}{Setup parameters} & $V_2=1.5$, $\lambda_2^2=2.5\%$ & $V_2^{(2)}=1.3$, $\lambda_2^2=5\%$ & $V_1^{(1)}=4.815$ \\
\multicolumn{1}{|c|}{} & $V_3=1.1$, $\lambda_3^2=2.5\%$ & & \\ \cline{1-4}
\multicolumn{1}{|c|}{Channel parameters} & T & $T^{(2)}\approx 0.999T$ & $T^{(1)}\approx 0.993T$ \\
\multicolumn{1}{|c|}{} & $\epsilon=0.05$ & $\epsilon^{(2)}\approx 0.0535$ & $\epsilon^{(1)}\approx 0.0773$ \\ \cline{1-4}
\end{tabular}
\caption{\label{equivtable}Effective parameters of the channel depending on the knowledge of trusted parties on the multimode structure of source and detectors.}
\end{table}
\end{center}
\end{widetext}
It is evident from the table, that as trusted parties have less information about the mode structure, the effective channel noise, which is present in the covariance matrix they measured, increases. This leads to the reduction of the lower bound on the key rate and to the limited applicability of the protocol, as illustrated in Fig. \ref{3scen}. Thus, the "ignorant" approach to the multimode detection may still provide security of CV QKD, but in the very limited range compared to the case, when the structure of the source and detectors is known and taken into account in the security analysis of the CV QKD protocols.
%%%%%%%%%%%%%%%%%%%%%%%%%%%%%%%%%%%%%%%%%%%%%%%%%%%%%%%%%%%%%%%%%%%%%%%%%%%%%%%%%%%%%%%%%%%
%%%%%%%%%%%%%%%%%%%%%%%%%%%%%%%%%%%%%%%%%%%%%%%%%%%%%%%%%%%%%%%%%%%%%%%%%%%%%%%%%%%%%%%%%%%%
The example given above illustrates the importance of knowledge of the mode structure.

{\em Symmetrization of the source modes.} In the case when the trusted parties are able to control the variance of the individual modes in the source, they can make it more balanced, which brings the untrusted homodyne detection scenario closer to the trusted one. If the source balancing is perfect, the single-mode scenario can be recovered. From (\ref{corrmat}) it is easy to see already for the two-mode case that if the source is perfectly balanced, i.e., $V_1 = V_2$, then $\sigma_{(A_1B_1),(A_2B_2)}=0$. This means that the correlation between the trusted modes and the untrusted outputs of the detection completely vanishes. In this regime the protocol becomes equivalent to the single-mode CV QKD protocol. Such positive effect of the source balancing is valid for any number of the modes. The multimode structure in this case effectively vanishes due to the averaging of the covariance matrix, which becomes fully equivalent to the single-mode one. Moreover, the energy leakage from the auxiliary modes of the homodyne detector stops in this case since their respective correlations with the signal modes are canceled out. In this case the difference between the trusted and untrusted homodyne detection vanishes, as can be seen from Fig. \ref{contour}, where contour lines are crossing the diagonal in the same points at both the left and right plots. The effect is similar to vanishing of information leakage from the lossy channel by state engineering \cite{Usenko2011} or to entanglement-induced transparency \cite{Olivares2009, Bloomer2011}. 
%%%%%%%%%%%%%%%%%%%%%%%%%%%%%%%%%%%%%%%%%%%%%%%%%%%%%%%%%%%%%%%%%%%%%%%%%%%%%%%%%%%%%%%%%%%
\begin{figure}
\centerline{\psfig{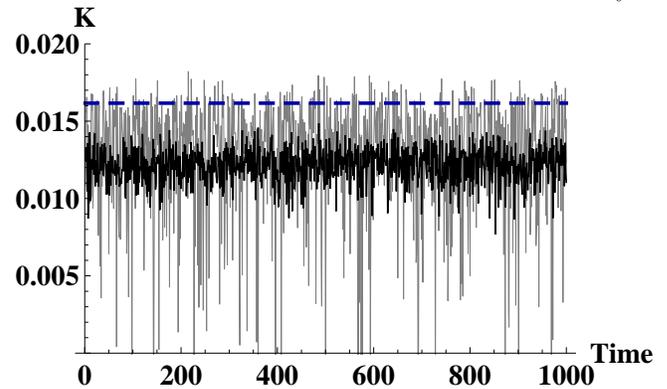}}
\caption{(Color online) Numerical simulation of lower bound on the secure key rate in case of collective attacks for multimode CV QKD with varying mode variance $V\sim\mathcal{N}(3,0.75)$ for number of modes $N=5$ (light gray) and $N=100$ (black) versus the time flow of the protocol runs. The single-mode CV QKD key rate for constant source variance $V=3$ is given in blue (dark gray) dashed line for the reference. In all the cases channel transmittance $T=3 \cdot 10^{-2}$ (corresponding to approximately 76 km of telecom fiber), channel noise $\epsilon=5\%$ SNU and postprocessing efficiency $\beta=95\%$.}
\label{KTime}
\end{figure}
%%%%%%%%%%%%%%%%%%%%%%%%%%%%%%%%%%%%%%%%%%%%%%%%%%%%%%%%%%%%%%%%%%%%%%%%%%%%%%%%%%%%%%%%%%%%

{\em Security stabilization.} Still, if the source is nearly balanced, but the asymmetries in the mode energy remain present, the key rate would be reduced. Surprisingly, such asymmetries or energy fluctuations per mode of the source have a less pronounced effect on the security of CV QKD when the number of modes is increased; this can be seen in Fig. \ref{KTime}. Our result constitutes the deterministic method of stabilization of the secure key rate, when errors are suppressed due to the joint homodyne detection of the many modes, used to encode information. The result opens a promising pathway towards CV QKD with the heavily multimode sources even upon energy fluctuations.

\section{Conclusions}
We suggest the CV QKD protocol using the multimode entangled states and homodyne detection. We show that multimode quantum key
distribution with homodyne detection is more feasible if the multimode structure is
symmetrically engineered. If the modes are unbalanced, we can expand the security region of the protocol
by taking into account the mode structure of the source. Moreover, we suggest the direct application of
multimode entanglement generation and detection, as more modes of the key
distribution stabilize fluctuations of the secure key rate. This result is in contrast to previous research, when the multimode character 
of quantum communication has been ignored or considered as negative. 
It motivates a search for the multimode quantum information
protocols where multimode structure is not limiting and even brings a positive effect. The result opens a pathway towards the experimental investigation and practical application of the multimode bright
entangled states in quantum communication.

%%%%%%%%%%%%%%%%%%%%%%%%%%%%%%%%%%%%%%%%%%%%%%%%%%%%%%%%%%%%%%%%%%%%%%%%%%%%%%%%%%%%%%%%%%
\medskip
\noindent {\bf Acknowledgments.} The research leading to these results has received funding from the EU FP7 under Grant Agreement No. 308803 (Project BRISQ2), co-financed by M\v SMT \v CR (7E13032), and has been partly supported by NATO SPS Project No. 984397. V.C.U. and L.R.  acknowledge the Project No. 13-27533J of GA\v CR.


\begin{thebibliography}{99}

\bibitem{Zeilinger2000}
{\em The Physics of Quantum Information}, edited by D. Bouwmeester, A. Ekert, and A. Zeilinger (Springer, Berlin, 2000).

\bibitem{Kok2010}
P. Kok and B. W. Lovett, Introduction to Optical Quantum Information Processing (Cambridge University Press, Cambridge, 2010).

\bibitem{Gisin2002}
N. Gisin, G. Ribordy, W. Tittel, and H. Zbinden, Rev. Mod. Phys. \textbf{74}, 145 (2002).

\bibitem{Scarani2009}
V. Scarani, H. Bechmann-Pasquinucci, N. J. Cerf, M. Du\v sek, N. Lutkenhaus, and M. Peev, Rev. Mod. Phys.
\textbf{81}, 1301 (2009).

\bibitem{Peres1993}
A. Peres, Quantum theory: concepts and methods (Kluwer, Dordrecht, 1993)

\bibitem{BB84}
C. H. Bennett, and G. Brassard, in Proceedings of the International Conference on Computers, Systems and Signal Processing (IEEE, Bangalore India, 1984), p. 175.

\bibitem{Ekert}
A. K. Ekert, Phys. Rev. Lett. \textbf{67}, 661 (1991).

\bibitem{Bell}
J. S. Bell, Physics \textbf{1}, 195 (1965).

\bibitem{Acin2007}
A. Ac\' in, N. Brunner, N. Gisin, S. Massar, S. Pironio, and V. Scarani, Phys. Rev. Lett. \textbf{98}, 230501 (2007).

\bibitem{Liu2013}
Y. Liu et al., Phys. Rev. Lett. \textbf{111}, 130502 (2013).

\bibitem{Ralph1999}
T. C. Ralph, Phys. Rev. A \textbf{61}, 010303 (1999).

\bibitem{Cerf2001}
N. J. Cerf, M. L\' evy, and G. Van Assche, Phys. Rev. A \textbf{63}, 052311 (2001).

\bibitem{Grosshans2002}
F. Grosshans and P. Grangier, Phys. Rev. Lett. \textbf{88}, 057902 (2002).

\bibitem{Grosshans2003a}
F. Grosshans, G. Van Assche, J. Wenger, R. Brouri, N.J. Cerf, and P. Grangier, Nature (London) \textbf{421}, 238 (2003).

\bibitem{Silberhorn2002}
Ch. Silberhorn, T. C. Ralph, N. L\" utkenhaus, and G. Leuchs, Phys. Rev. Lett. \textbf{89}, 167901 (2002).

\bibitem{Jouguet2013}
P. Jouguet,	 S. Kunz-Jacques, A. Leverrier, P. Grangier, and E. Diamanti, Nat. Photon. \textbf{7}, 378–381 (2013).

\bibitem{Patron2009}
R. Garc\' ia-Patron and N. J. Cerf, Phys. Rev. Lett. \textbf{102}, 130501 (2009).

\bibitem{Filip2008}
R. Filip, Phys. Rev. A \textbf{77}, 022310 (2008).

\bibitem{Usenko2010}
V. C. Usenko and R. Filip, Phys. Rev. A \textbf{81}, 022318 (2010).

\bibitem{Weed2010}
Ch. Weedbrook, S. Pirandola, S. Lloyd, and T. C. Ralph, Phys. Rev. Lett. \textbf{105}, 110501 (2010)

\bibitem{Weed2012}
Ch. Weedbrook, S. Pirandola, and T. C. Ralph, Phys. Rev. A \textbf{86}, 022318 (2012)

\bibitem{Usenko2011}
V. C. Usenko and R. Filip, New J. Phys. \textbf{13}, 113007 (2011).

\bibitem{Madsen2012}
L. S. Madsen, V. C. Usenko, M. Lassen, R. Filip, and U. L. Andersen, Nat. Commun. \textbf{3}, 1083 (2012).

\bibitem{Ou1992}
Z. Y. Ou, S. F. Pereira, H. J. Kimble, and K. C. Peng, Phys. Rev. Lett. \textbf{68}, 3663 (1992).

\bibitem{Opatrny2002}
T. Opatrn\' y, N. Korolkova, and G. Leuchs, Phys. Rev. A \textbf{66}, 053813 (2002)

\bibitem{Wasil2006}
W. Wasilewski, A. I. Lvovsky, K. Banaszek, and C. Radzewicz, Phys. Rev. A \textbf{73}, 063819 (2006).

\bibitem{Christ2011}
A. Christ, K. Laiho, A. Eckstein, and K. N. Cassemiro, Ch. Silberhorn, New J. Phys. \textbf{13} 033027 (2011).

\bibitem{Hockel2011}
D. H\"ockel, L. Koch, and O. Benson, Phys. Rev. A \textbf{83}, 013802 (2011).

\bibitem{Poly2012}
C. Polycarpou, K. N. Cassemiro, G. Venturi, A. Zavatta, and M. Bellini, Phys. Rev. Lett. \textbf{109}, 053602 (2012).

\bibitem{Morin2013}
O. Morin, C. Fabre, and J. Laurat, Phys. Rev. Lett. \textbf{111}, 213602 (2013).

\bibitem{Helwig2009}
W. Helwig, W. Mauerer, and Ch. Silberhorn, Phys. Rev. A \textbf{80}, 052326 (2009)

\bibitem{Perina2007}
J. Perina, J. K\v repelka, J. Pe\v rina Jr, M. Bondani, A. Allevi, and A. Andreoni, Phys. Rev. A \textbf{76} 043806 (2007).

\bibitem{Isk2009}
T. Sh. Iskhakov, M. V. Chekhova, and G. Leuchs, Phys. Rev. Lett. \textbf{102}, 183602 (2009).

\bibitem{Agafonov2010}
I. N. Agafonov, M. V. Chekhova, and G. Leuchs, Phys. Rev. A \textbf{82}, 011801(R) (2010).

\bibitem{Allevi2012}
A. Allevi, S. Olivares, and M. Bondani, Phys. Rev. A \textbf{85}, 063835 (2012).

\bibitem{Perina2012}
J. Pe\v rina Jr., M. Hamar, V. Mich\' alek, and O. Haderka, Phys. Rev. A \textbf{85}, 023816 (2012).

\bibitem{Isk2012}
T. Sh. Iskhakov, I. N. Agafonov, M. V. Chekhova, and G. Leuchs, Phys. Rev. Lett. \textbf{109}, 150502 (2012).

\bibitem{Brida2010}
G. Brida, M. Genovese, and I. R. Berchera, Nat. Photon. \textbf{4}, 227 (2010).

\bibitem{Brida2011}
G. Brida, M. Genovese, A. Meda, and I. R. Berchera, Phys. Rev. A \textbf{83}, 033811 (2011).

\bibitem{Lopaeva2013}
E. D. Lopaeva, I. Ruo Berchera, I. P. Degiovanni, S. Olivares, G. Brida, and M. Genovese, Phys. Rev. Lett. \textbf{110}, 153603 (2013).

\bibitem{Opatrny1997}
T. Opatrny, D.-G. Welsch, and W. Vogel, Phys. Rev. A \textbf{55}, 1416 (1997).

\bibitem{McAlister1997}
D. F. McAlister and M. G. Raymer, Phys. Rev. A \textbf{55}, R1609 (1997).

\bibitem{Fiurasek2001}
J. Fiur\' a\v sek, Phys. Rev. A \textbf{63}, 033806 (2001).

\bibitem{twinbeams}M. Vasilyev, S.-K. Choi, P. Kumar, and G. Mauro D'Ariano, Phys. Rev. Lett. \textbf{84}, 2354 (2000).

\bibitem{cktheorem}I. Csisz\'ar, and J. K\"orner, IEEE Trans. Inf. Theory \textbf{24} (3), 339 (1978).

\bibitem{generalattacks1}R. Renner, Nat. Phys. \textbf{3}, 645 (2007).

\bibitem{generalattacks2}M. Mertz, H. Kampermann, S. Bratzik, and D. Bru\ss , Phys. Rev. A \textbf{87}, 012315 (2013).

\bibitem{cvcohatt}A. Leverrier, R. Garc\'ia-Patr\'on, R. Renner, and N. J. Cerf, Phys. Rev. Lett. \textbf{110}, 030502 (2013).

\bibitem{Weed2013}
Ch. Weedbrook, Phys. Rev. A \textbf{87}, 022308 (2013).

\bibitem{lodewyck}J. Lodewyck, et al., Phys. Rev. A \textbf{76}, 042305 (2007).

\bibitem{effectivecodes}P. Jouguet, S. Kunz-Jacques, and A. Leverrier, Phys. Rev. A \textbf{84}, 062317 (2011).

\bibitem{extremality}M. M. Wolf, G. Giedke, and J. I. Cirac, Phys. Rev. Lett. \textbf{96}, 080502 (2006).

\bibitem{optimality1}M. Navascues, F. Grosshans, and A. Ac\'in, Phys. Rev. Lett. \textbf{97}, 190502 (2006).

\bibitem{optimality2}R. Garc\'ia-Patr\'on, and N. J. Cerf, Phys. Rev. Lett. \textbf{97}, 190503 (2006).

\bibitem{entrfunct}A. Serafini, M. G. A. Paris, F. Illuminati, and S. De Siena, J. Opt. B: Quantum Semiclass. Opt. \textbf{7}, R19 (2005).

\bibitem{Weed2012b}
Ch. Weedbrook, S. Pirandola, R. Garcia-Patron, N. J. Cerf, T. C. Ralph, J. H. Shapiro, and S. Lloyd, Rev. Mod. Phys. \textbf{84}, 621 (2012).

\bibitem{Olivares2009}
S. Olivares and M. G. A. Paris, Phys. Rev. A. \textbf{80}, 032329 (2009).

\bibitem{Bloomer2011}
R. Bloomer, M. Pysher, and O. Pfister, New J. Phys. \textbf{13}, 063014 (2011)


\end{thebibliography}
\end{document}